# Pressure induced superconductivity in a CeRhSi$_3$ single crystal – the high pressure study

Daniel Staško[1,*], Jaroslav Valenta[1,2], Marie Kratochvílová[1], Jiří Prchal[1], Petr Proschek[1], Milan Klicpera[1,*]

[1]*Charles University, Faculty of Mathematics and Physics, Department of Condensed Matter Physics, Ke Karlovu 5, 12116 Prague 2, Czech Republic*
[2]*National Institute for Materials Science, Thermal Energy Materials Group, International Center for Materials Nanoarchitectonics (MANA), 1-2-1, Sengen, Tsukuba, Ibaraki 305-0047, Japan*

*Corresponding authors: staskodaniel@gmail.com, mi.klicpera@seznam.cz*

**Abstract**

Pressure induced superconductivity in non-centrosymmetric CeRhSi$_3$ and CeIrSi$_3$ compounds has attracted significant attention of the scientific community since its discovery 15 years ago. Up-to-date, all reported experimental results were obtained employing the hybrid-cylinder piston pressure cells with a maximum reachable pressure of 3 GPa. Present study focuses on the superconducting state at higher, so far unreported, pressures using the Bridgman anvil cell and a CeRhSi$_3$ single crystal synthesized by the Sn-true-flux method. The initial increase of superconducting critical temperature from 0.4 K at 1.1 GPa to 1.1 K at 2.4 GPa is followed by a gradual suppression of SC state upon increasing the pressure above 3.0 GPa, forming a typical dome. The pressure induced superconductivity is expected to be completely suppressed in the pressure region between 4.5 and 5.0 GPa. Temperature dependence of electrical resistivity in constant magnetic fields and high pressures, as well as the magnetoresistance measurements, reveal a large critical field, exceeding 19 T at 0.6 K and 2.4 GPa, sharply decreasing receding the superconductivity dome. The previously reported *T-p* and *H-T* phase diagrams are completed by our high-pressure data and discussed in the frame of previous results.

# 1. Introduction

Ce-based intermetallic compounds frequently exhibit complex magnetic and transport properties connected to a single electron localized at 4f-level and an interplay among indirect exchange interactions, Kondo screening, spin-orbit coupling, and crystal electrical field (CEF). A family of the Ce$TX_3$ ($T$ – transition d-metal, $X$ – p-metal) compounds has been reported to host a variety of exotic properties: Ce valence fluctuations in CeRuSi$_3$ evidenced by inelastic neutron scattering (INS) [1,2]; spin-glass state disclosed in CePtAl$_3$ [3]; complex magnetic ordering with several magnetic phase transitions and rich phase diagram in CePtSi$_3$ [4] and CeCoGe$_3$ [5,6], which, moreover, becomes superconducting (SC) under applied hydrostatic pressure [7,8]. The pressure induced superconductivity was evidenced also in the non-centrosymmetric analogues CeRhSi$_3$ [9] and CeIrSi$_3$ [10]. A connection among magnetism, superconductivity, and valence fluctuations was discussed in the heavy fermion (HF) antiferromagnet CeRhGe$_3$ [11,12,13]. Heavy-fermion behaviour of CeCuAl$_3$ [14,15] was ruled out considering the low-energy CEF excitation observed in INS spectra [16]. Simultaneously, INS data contained additional excitation ascribed to a magneto-elastic quasi-bound state [16,17,18,19]. Similar spectra were recently reported for CeAuAl$_3$, containing additional CEF-phonon spectral crossing [20]. For above mentioned physical properties Ce$TX_3$ have remained of a scientific interest to the condensed matter physics community.

CeRhSi$_3$, the subject of present study, as well as other Ce$TX_3$ compounds, crystallizes in an ordered tetragonal structure of the BaNiSn$_3$-type ($I\,4\,m\,m$, no. 107). Tetragonal planes of respective elements are stacked along the c-axis, Ce-$T$-$X$-$X$-Ce, creating a non-centrosymmetric lattice. The materials without the inversion centre were believed not to be consistent with a superconducting state – spin-triplet pairing requires a parity of unit cell. Indeed, most of HF superconductors crystallize in a centrosymmetric crystal structure [21]. However, several superconductors adopting a lattice lacking the inversion symmetry have been reported besides CeRhSi$_3$ [9]: CePt$_3$Si [22] and CeIrSi$_3$ [10]. The superconductivity in these materials cannot be explained by a standard singlet-triplet pairing, as spin and orbital contributions cannot be treated independently [9,23]. It is believed that the spin-orbit-coupling (SOC), specifically Rashba-type antisymmetric SOC [24], together with a strongly fluctuating order parameter, play a significant role in formation of superconductivity in these materials.

Large γ coefficient ($\gamma$ = 120 mJ.mol$^{-1}$.K$^{-2}$, similar for CeIrSi$_3$), compared to other analogues [12], lists CeRhSi$_3$ among heavy-fermion materials. A competition between RKKY and Kondo interactions, together with almost itinerant 4f electrons, stand behind the small magnetic moment of Ce ions in CeRhSi$_3$ [25,26]. The antiferromagnetic structure bellow 1.6 K is described by the incommensurate propagation vector (~0.215, 0, 1/2); the magnetic moments, of the order of 0.1 μ$_B$/Ce, form the spin-density-wave structure [12,25]. Recently, the antiferromagnetic order in CeRhSi$_3$, based on the neutron spectroscopy results [27], was suggested to be driven by the weakly correlated relaxing Kramers' doublets. Upon the pressure application, the Néel temperature first slightly increases up to 2 K at 1.0 GPa and subsequently decreases down to 1.1 K at 2.4 GPa where the magnetic transition coincides with the superconducting transition (at T$_c$) and cannot be further distinguished in the electrical resistivity data [23,28,29]. Pressure induced SC emerges at ~1.2 GPa for electrical current, $j$, applied along [100] crystallographic direction, while the measurement with $j \parallel$ [001] reveals the SC transition at a significantly lower pressure [23]. The emergence of SC transition is quite continuous and is accompanied by an additional transition at T$^*$ in $j \parallel$ [100] data [9,23]. The T$^*$ anomaly was reproduced also on the temperature evolution of ac-susceptibility measured with magnetic field applied along [100]. Both types of measurements along [001] revealed a single SC transition [9,23]. T$_c$ forms a plateau at around 2.6 GPa and is expected to decrease at higher pressures. Measurements in magnetic field revealed a huge upper critical field, possibly reaching 30 T with $H \parallel$ [001], while an upper limit of 8 T for $H \parallel$ [100] was reported [31]. A significant anisotropy in SC's evolution was attributed to the absence of the paramagnetic pair-breaking effect. Furthermore, pressure dependence of the critical field was studied in detail to investigate the interplay between magnetism and superconductivity [28,30,32] as well as a possible quantum critical point [33,34,35]. The quantum critical point inside the superconductivity dome was discussed also in our recent study on Ge-doped CeRh(Si,Ge)$_3$ under hydrostatic pressure [36]. The penetration depth measurements indicated that the magnetism and superconductivity coexist up to high pressures [29].

Previous high-pressure experiments on CeRhSi$_3$ were performed utilizing the hybrid-cylinder piston pressure cell with a maximum reachable pressure of 3 GPa. The measurement at pressures higher than 3 GPa was missing, and is the subject of our present study. We employed the Bridgman anvil cell with liquid pressure medium to reach the pressure as high as 4.3 GPa. The measured data allowed to complete the previously reported T-p diagram [23,28,29,37,38], as well as to follow the superconductivity in external magnetic field up to higher pressure. The acquired results are discussed with respect to previous reports on CeRhSi$_3$ and other Ce$TX_3$ pressure induced superconductors.

## 2. Sample preparation and characterization, experimental methods

CeRhSi$_3$ single crystals were reported to be prepared employing Czochralski method and arc-furnace [9,29,39]. We attempted to prepare the single crystal following the reported preparation route, repeatedly, leading to multi-phase samples, containing, besides CeRhSi$_3$ phase, also CeRh$_2$Si$_2$ and binary phases. Therefore, the further investigated single crystal was synthesized using the true-flux method with Sn being the solvent. Usually, the flux method allows to prepare relatively small single crystals, whose dimensions are, nevertheless, sufficient (and in fact, preferable) for high-pressure experiments. A number of crystals with dimensions of ~(100 x 250 x 900) μm$^3$ were synthesized, see Fig.1a. Composition and homogeneity of the samples' surface were investigated by the scanning electron microscope MIRA (Tescan) equipped with back-scattered electron (BSE) and energy dispersive X-ray (EDX) detectors. The stoichiometry Ce:Rh:Si was determined to be 0.95(9):1.05(9):3.00(9). No secondary phases were observed. The perfect quality of single-crystalline samples was further confirmed employing Laue X-ray diffraction (Fig.1b).

Contrary to relatively large Czochralski grown single crystals, which have to be carefully oriented and cut to prepare the samples suitable for high-pressure experiments, the crystals synthesized by the flux method have several practical advantages. The small single crystals are of very good quality, homogeneity, and crystallinity, and importantly, they grow with prominent facets along/perpendicular to principal crystallographic directions. We highlight the top-right crystal in Fig.1a, which grew along two perpendicular directions, suggesting the 4-fold symmetry of the third perpendicular direction. Indeed, Laue diffraction unambiguously identified the [001] direction to be perpendicular to the samples' surface (Fig.1b), i.e. the c axis coincides with the thinnest dimension of the crystals. Somewhat surprisingly the long edge of the crystals was not determined to be [100], but the [110] direction instead. On one hand, the experimental setup – arrangement within the pressure cell – is strongly limited by the sample shape and dimensions, i.e. the electrical current can be reasonably applied only along the longest edge. On the other hand, the electrical current and magnetic field are applied precisely along individual axes. The facets of the flux-grown crystals are naturally oriented and the samples are sufficiently small – the smallest dimension, along [001], was reduced by polishing to 80 μm to fit into the pressure cell, as well as to completely remove the Sn residue at the samples' surface. No sample cutting was necessary, preventing any, even slight, mis-orientation.

Specific heat and electrical resistivity at ambient pressure were measured as another part of the characterization routine. The heat capacity was measured down to 0.4 K employing Physical Property Measurement System (PPMS, Quantum Design) and a standard time-relaxation method on a sample of 0.14 mg mass. PPMS and 4-probe method were used for electrical resistivity, $\rho(T)$, and magnetoresistance, $\rho(H)$, measurements. Temperature developments of both quantities are well in agreement with previous results [23,39,40] and demonstrate, at least, comparable quality of investigated crystals. Specific heat data revealed a pronounced lambda anomaly at $T_N$ = 1.6 K, representing the transition from paramagnetic to antiferromagnetic state (Fig.1c). The well pronounced anomaly, compared to previous reports [40], can be ascribed to the high quality of the investigated single crystal. Magnetic field up to 12 T applied along [001] affected the phase transition only slightly, shifting the magnetic entropy related to the transition to the higher temperature. The electrical resistivity data exhibited a pronounced kink at $T_N$ (Fig.1d), consistently with the specific heat measurement. Although the electrical current was applied along the [110] direction, while the previous studies were performed with $j$ along [100] or [001] [23,39,40], the development of measured electrical resistivity – including its magnitude and the so called RRR factor, RRR = $R_{300K}/R_{0.4K}$ = 88 – mimicked the reported results, especially for $j \parallel$ [100].

Bridgman anvil cell (BAC; Fig.1e) and a liquid pressure transmitting medium, Daphne Oil 7373, were used for the high-pressure experiment on $CeRhSi_3$. BAC with liquid medium provides more hydrostatic conditions than BAC with solid exchange medium, although the Daphne 7373 solidifies at 2.2 GPa at room temperature [41]. With a theoretical limit of 6 GPa, this pressure cell was employed to measure resistivity with the AC four probe method. Alongside the sample, a lead manometer was installed within the sample space (Fig.1e,f) to determine the actual pressure at low temperatures. The electrical resistivity was measured by 12 insulated Cu wires (25 μm in diameter) bonded to the sample/lead, fixed by stycast epoxy around the support cone and led through the middle of bottom anvil. Due to high probability of wire/s loss during the experiment, 2 spare wires were used for each – sample and lead. Two identical pressure cells were prepared for the measurements, allowing both to confirm a reproducibility of the measurement and to sufficiently cover especially the high-pressure region of the $CeRhSi_3$ phase diagram. The experiment was conducted at temperatures down to 0.3 K and in magnetic fields up to 19 T employing '20 T & 30mK' system, Cryogenic Limited. The pressure was tuned off-line at room temperature. The pressure increase was estimated empirically by load calibration, and measured precisely at low temperature utilizing the knowledge on a pressure dependence of superconductivity of lead [42].

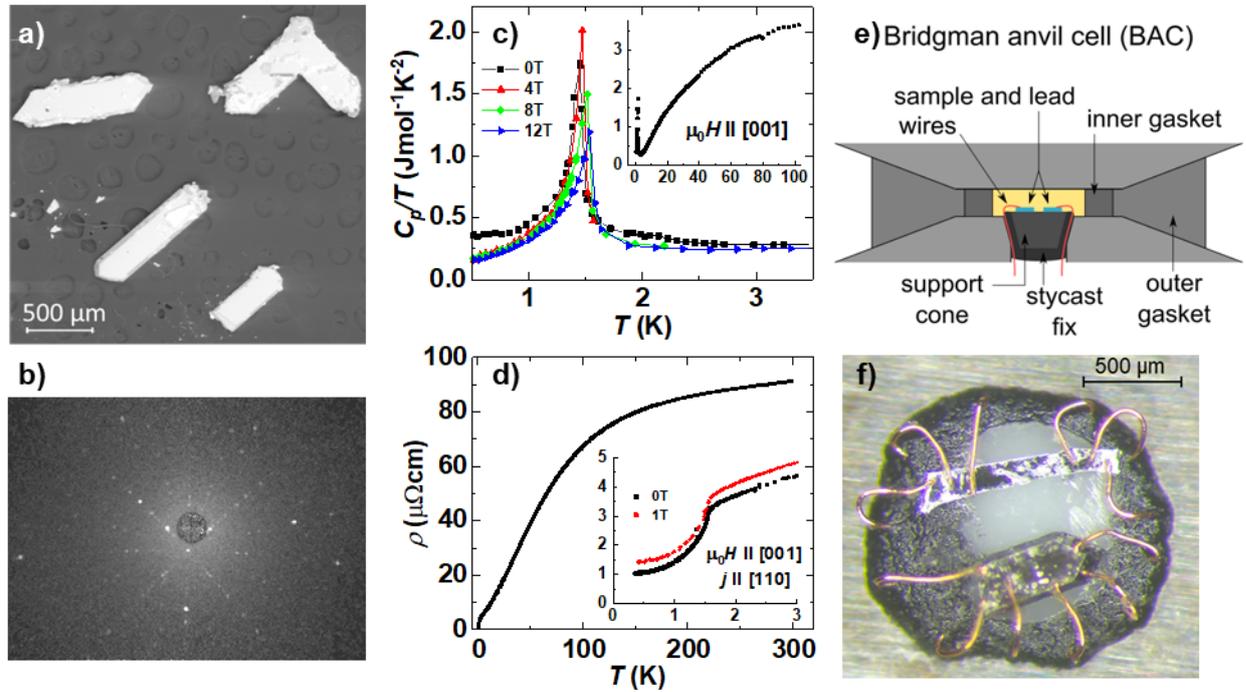

Fig.1: Single crystal characterization and pressure cell setup: a) BSE image of several as-grown $CeRhSi_3$ single crystals, Sn-flux on the surface was later removed by polishing; b) Laue X-ray diffraction patterns taken along [001]; c) specific heat measurement with zoomed low-temperature lambda-type anomaly; d) electrical resistivity measurement at ambient pressure; e) scheme of the BAC sample space; and f) top view of the experimental montage of $CeRhSi_3$ (bottom) and lead (top) within the sample space.

## 3. Experimental results

First, let us describe the resistivity in the $T$-region between 300 K and 2 K at pressure of 0.3 GPa. The electrical resistivity exhibits a similar temperature dependence as observed at ambient pressure, compare Figs.1d and 2a. Following $\rho(T)$ on cooling from the room temperature, a subtle decrease of resistivity becomes notably steeper below ~150 K. Applying larger pressure results in a steeper decrease of overall resistivity, while room temperature value of ρ(T) is higher. Such development is ascribed to the changes of crystallographic lattice under pressure. The applied pressure influences both interatomic distances and local environment of individual ions, leading to changes of the phonon as well as magnetic contributions to electrical resistivity.

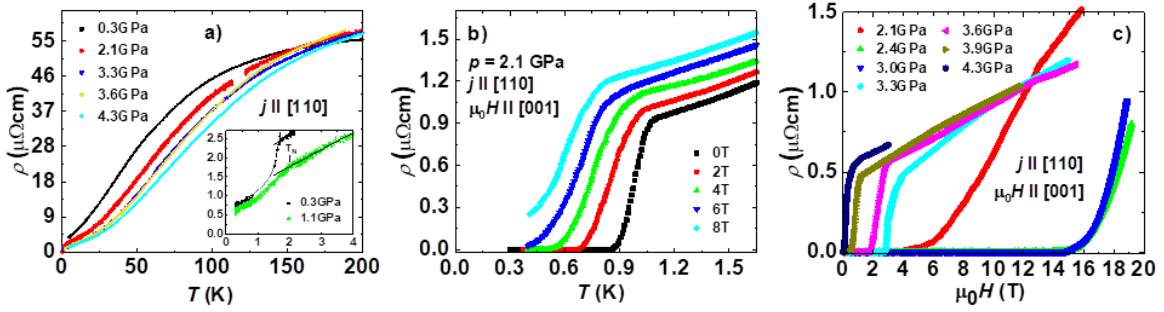

Fig.2: Transport properties measurements on a CeRhSi$_3$ single crystal under pressure: a) temperature dependence of electrical resistivity below 200 K, zoomed low-temperature region displays the kink related to the AFM phase transition. The ordering temperature was determined fitting the electrical resistivity above the resistivity drop by linear function and observing the deviation of data from the fit; b) temperature dependence of resistivity at 2.1 GPa in magnetic field applied along [001]; c) magnetoresistance measurements at lowest achievable temperature $T$ = 0.3 K and high pressures.

In further text, we focus on the low-temperature region of measured data where anomalies corresponding to the AFM phase transition and SC transition are observed (Fig.3). CeRhSi$_3$ was reported to order antiferromagnetically below $T_N$ = 1.6 K [9,25], corresponding to the kink in our ambient pressure $\rho(T)$, Fig.1d. A similar resistivity development was followed under pressure of 0.3 GPa with $T_N$ slightly shifted to higher temperature (1.65(5) K), see Fig.2a. Increasing the pressure up to 1.1 GPa, the transition becomes less pronounced and shifts to even higher temperature (2.00(9) K). No anomaly possibly ascribed to the antiferromagnetic transition is observed in our 1.8 GPa data. The sensitivity of experimental setup showed to be insufficient to unambiguously determine very small kinks in temperature dependencies of electrical resistivity, according to previous results expected at 1.8 and 2.1 GPa [9,23]. Alternatively, the AFM transition coincides with SC transition already at 1.8 GPa. No anomaly related to AFM is detected at higher pressures, well in agreement with previous results [9,23,28,29,30,31,32,33], see Fig.4.

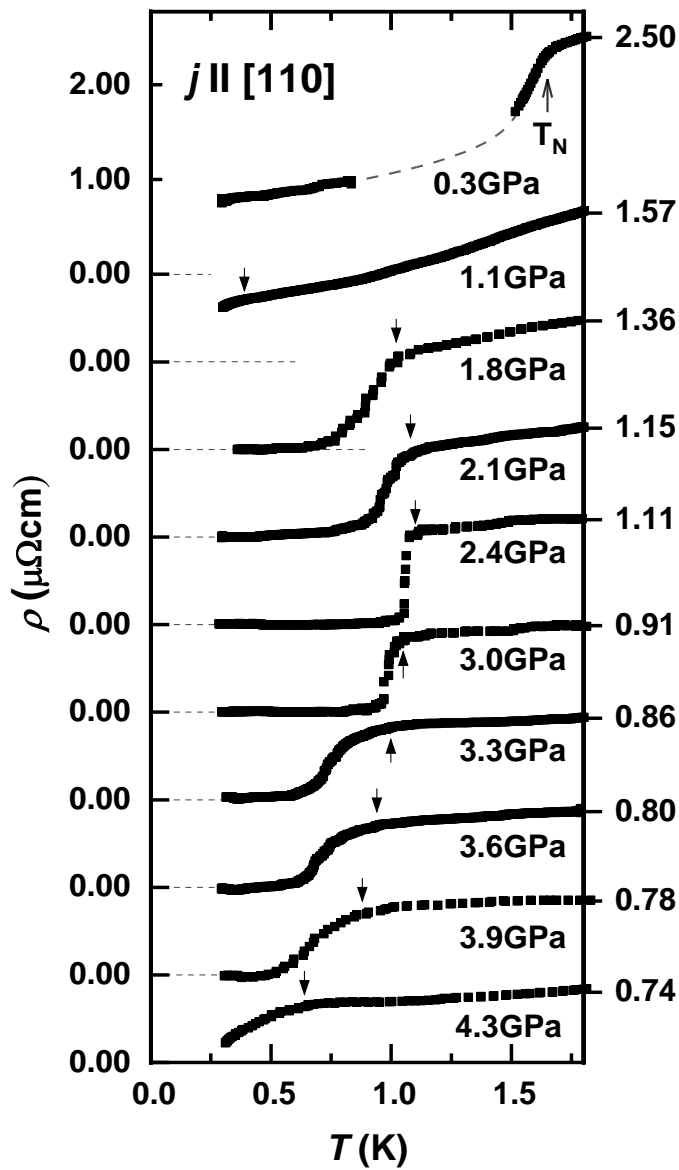

Fig.3: Electrical resistivity data with pronounced magnetic and superconducting transitions under applied pressure. Left and right axes are chosen to best demonstrate the evolution of electrical resistivity in individual pressures. The arrows indicate $T_c$, except the arrow for 0.3 GPa data marking $T_N$.

Turning our attention to pressure induced superconductivity in CeRhSi$_3$, the SC state is not fully reached down to 0.3 K at 1.1 GPa, however an unusual decrease in resistivity is observed when approaching the lowest temperatures (instrumental limits), see Fig.3. Following previous studies [9,23], the critical SC temperature, $T_c$, is determined as a high-temperature onset of the electrical resistivity drop/decrease (Fig.3). Determined $T_c$ agrees well with previous results [9,23] measured with $j \parallel [100]$ and is used, together with the higher-pressure data, to complete the *T-p* diagram in Fig.4. Critical temperature of a pronounced SC transition at 1.8 GPa fits perfectly to the reported phase diagram. A maximal value of $T_c = 1.10(4)$ K is observed at 2.4 GPa. Also, a decrease of the electrical resistivity during the SC transition is sharper compared to the lower- and higher-pressure data. At 3.0 GPa, SC transition is still well pronounced, but shows the first signs of SC suppression, having slightly smaller $T_c$ and being broader than at 2.4 GPa. Further pressure application resulted in a continuous suppression of SC, shifting $T_c$ to lower temperatures, and broadening the SC transition. Reaching 4.3 GPa, the SC transition is almost suppressed, the SC state is not completely reached down to 0.3 K (Fig.3). Further increase of the pressure is expected to completely close the SC dome, at temperatures beyond our instrumental limits.

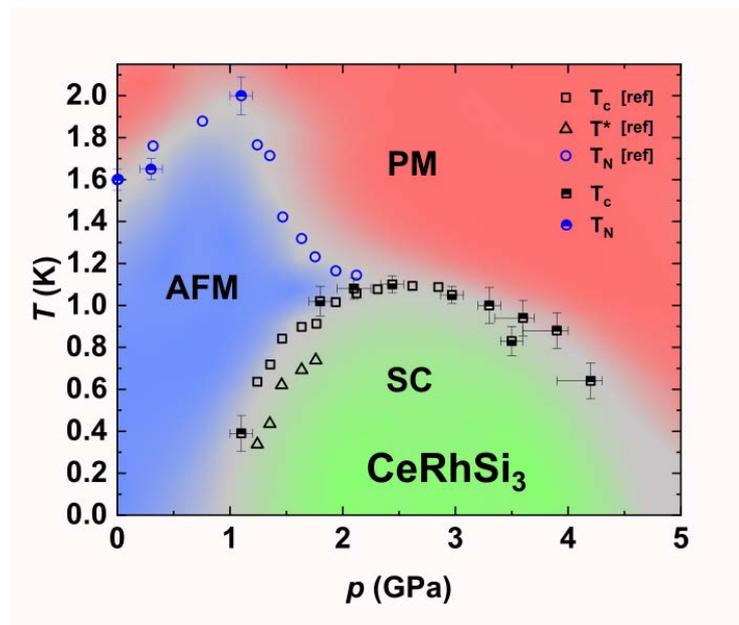

Fig.4: *T-p* phase diagram of pressure induced SC in CeRhSi$_3$. The data collected employing first (upper-part-filled symbols) and second pressure cell (bottom-part-filled symbols) are summarized in the plot. The data adopted from Ref. [23] were measured with $j \parallel [100]$ and are shown as empty symbols. See text for more details.3

Previously reported anomalous behaviour of the SC transition at T*, observed in electrical resistivity measurements for $j \parallel [100]$ between 1.2 and 1.8 GPa [9,23], is not reproduced in our data. There is no indication of additional/subsequent transition at high pressures, albeit the transitions are relatively broad. Although we cannot completely rule out the presence of the T* transition in our data, we tend to consider it to be sample dependent. Previously, the measurement was done on Czochralski grown single crystal, while the present study is performed on the flux-grown sample. Simultaneously, the present study is performed with $j \parallel [110]$, while previous measurements were done with current applied along [100]. No T* transition was observed measuring in the $j \parallel [001]$ arrangement [23].

Application of an external magnetic field, $H \parallel [001]$, suppresses the SC transition down to lower temperatures, as expected, in the whole pressure region. $\rho(T)$ development in static fields for 2.1 GPa is presented in Fig.2b as an example. Relatively sharp transition in zero field broadens continuously with increasing magnetic field. A similar field evolution of SC transition is observed for all the pressures. A large upper critical field, $H_c$, exceeding 19 T at ~0.6 K for both 2.4 GPa and 3.0 GPa is well in agreement with previously reported results [31], where the upper critical field was estimated to reach more than 30 T at zero temperature. Linearly extrapolating our data the critical field at zero temperature would reach value of 42 T. Receding from the SC dome apex the critical field decreases considerably, down to the value of 0.5 T at 4.3 GPa (see *H-T* diagram in Fig.5).

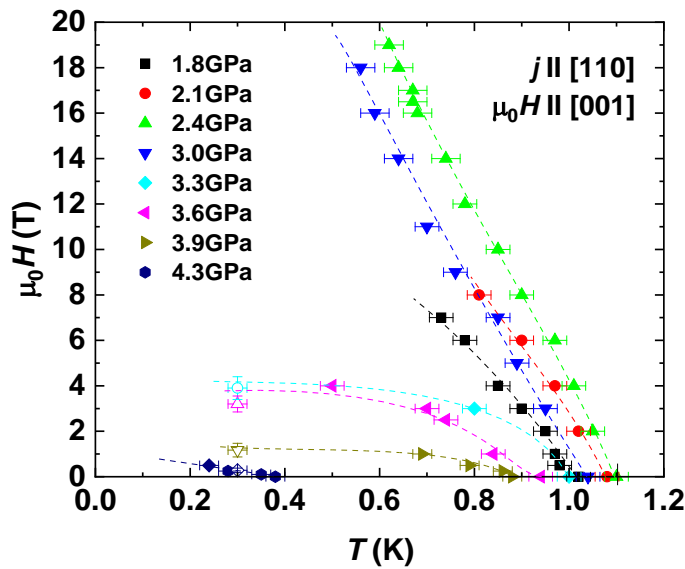

Fig.5: *H-T* phase diagram constructed from temperature (full symbol) and field (open symbols) dependencies of the electrical resistivity under pressure. The lines are guide to the eye.

The $\rho(T)$ data in constant magnetic field were complemented by magnetoresistance measurements. $\rho(H)$ curves measured at lowest achievable temperature for individual pressures are presented in Fig.2c. The SC transition first moves to higher fields applying external pressure up to 2.4 GPa, remains almost unchanged up to 3.0 GPa, and with a further application of pressure shifts to low fields. We highlight a significant changeover between SC transition under 3.0 GPa and all higher pressures; the critical field is drastically suppressed between 3.0 and 3.3 GPa, i.e. right after leaving the top of the SC dome (Fig.4). Any additional anomaly, previously introduced as $H^*$ [23], is not followed in our data, recalling the sample-dependent scenario.

*H-T* phase diagram in Fig.5 depicts the evolution of SC in magnetic field for all the pressures. Critical field is increasing with pressure from 1.8 GPa to 2.4 GPa, where SC transition is clearly observable even in 19 T (at ~0.6 K). Similarly robust critical field survives up to 3.0 GPa. Severely supressed critical field, unlike the moderately suppressed critical temperature, is observed receding the dome apex. The formation, as well as suppression, of the SC dome is reproduced in both critical temperature and critical field evolution with pressure, hence documenting a full consistence of the completed *H-T* diagram and *T-p* diagram in Figs.4 and 5.

## 4. Discussion

Gradual suppression of the pressure induced superconductivity above 3.0 GPa is characterized by a decrease of the critical temperature (Fig.3), and also by a significant broadening of the SC transition. Broad SC transitions are observed at lower pressures up to 2.1 GPa, see also previous results [9,30]. This effect is arguably connected with the sense of passing the SC boundary in the phase diagram. The non-zero width anomaly appears as broader one if it is being measured crossing the phase boundary not perpendicularly in the sense of the phase diagram – at the top of the SC dome between 2.4 and 3.0 GPa the SC transitions are sharper. Broadening of the SC transition can also be partly ascribed to non-ideal hydrostatic conditions at higher pressures. Used liquid transmitting medium solidifies at 2.2 GPa, creating additional shear stress at higher pressures, which can even result in a loss of contacts on the sample/lead. While this effect is negligible below 2.2 GPa (where broad transitions are nevertheless observed), it can considerably affect the measurements at higher pressures.

A significant anisotropy between individual electrical current directions, comparing previous data measured with $j \parallel [100]$ and $j \parallel [001]$ [9,23] to our data ($j \parallel [110]$), is revealed. Besides considerably different $T_c$'s (in lower-pressure region) for $j \parallel [001]$ and $j \perp [001]$ [9,23], the SC transition measured with $j \perp [001]$ was broader (Actually, second transition at $T^*$ was reported for $j \parallel [100]$.) compared to $j \parallel [001]$ direction [28,30,32]. Overall behaviour of the SC phase, as investigated in present paper, is in line with the previous results measured with $j \parallel [100]$ rather than those with $j \parallel [001]$, except the upper critical field of the value closer to latter direction. The anisotropy of the SC state is well understood considering the tetragonal lattice planes stacked along the [001] direction, and non-BCS nature of superconductivity in non-centrosymmetric $CeRhSi_3$. Noticeably similar properties were previously reported for isostructural analogue $CeIrSi_3$ [43] as well as for $CePt_3Si$ [22]. All the heavy-fermion compounds crystallize in a non-centrosymmetric tetragonal structure, order antiferromagnetically at low temperatures (below 5 K and 2.2 K, respectively), and reveal an unconventional superconductivity under pressure and at ambient pressure, respectively [10,22,43]. A broad transition, or even additional transition below $T_c$, at non-zero pressure, and anisotropic properties of SC state were reported for the three compounds [23,43,44]. The lack of inversion symmetry implies a lifting of the twofold spin degeneracy by a spin-orbit coupling, i.e. singlet and triplet pairings mixing, leading to a spin magnetic susceptibility anisotropy. To fully explain the unconventional SC and the anisotropy in tetragonal non-centrosymmetric compounds, the antiferromagnetic order should be considered as well [45,29,46]. A staggered moment acts on the SC gap; AFM could induce multiple SC transitions, as well as give rise to anisotropic behaviour.

Considering very similar structural, magnetic, and superconducting (under pressure) properties of $CeRhSi_3$ and $CeIrSi_3$ [23,29,47], and the phase diagrams constructed for former analogue (Figs.4 and 5), we are allowed to expect that (especially) the new high-pressure parts of diagrams are likely to be reproduced in $CeRhSi_3$, being investigated up to high pressures. Following the critical magnetic field in $CeRhSi_3$, a strong suppression of $H_c$ with applied pressure is observed receding the SC dome apex beyond 3.0 GPa. The large critical field, and its anisotropy, and its development with applied pressure were previously discussed in the frame of strong-coupling model, reduced g-factor, reduced paramagnetic pair-breaking mechanism (SOC), and helical vortex state [23 and references therein], similarly as in the case of CePt$_3$Si [45,48]. Although the pair-breaking mechanism was assumed to be most probably responsible for the observed behaviour, other interactions should be considered as well. Indeed, recent study [29] on both analogues, CeRhSi$_3$ and CeIrSi$_3$, highlighted the influence of antiferromagnetic order on Cooper pairs, and magnetic fluctuations associated with quantum-critical-point at $p_c$ = 2.8 GPa, i.e. at the SC dome apex, on superconductivity in those compounds. To fully understand the SC state in non-centrosymmetric compounds, and its pressure and magnetic field evolution, further studies, both experimental (namely e.g. synchrotron radiation and neutron scattering) and mainly theoretical, are highly desirable.

## 5. Conclusions

The pressure induced superconductivity in a flux-synthesized CeRhSi$_3$ single crystal was investigated up to 4.3 GPa employing the Bridgman anvil cell and transmitting medium Daphne oil 7373. The high-pressure experimental data were utilized to complete the CeRhSi$_3$ phase diagram sketched by the previously reported *T-p* and *H-T* dependencies. The signs of SC transition were observed at 1.1 GPa with $T_c$ = 0.4 K. Upon increasing pressure, the critical temperature increased up to 1.1 K at 2.4 GPa, and then beyond 3.0 GPa gradually decreased with further pressure application. A characteristic superconductivity dome was estimated to close between 4.5 and 5.0 GPa. Application of magnetic field shifted the SC transition to lower temperatures at all the pressures. The high critical field exceeding 19 T at 0.6 K and 2.4 and 3.0 GPa was significantly suppressed receding from the SC dome apex, consistently with the *T-p* phase diagram.

**Acknowledgement**

We thank to Petr Čermák for his assistance with '20 T & 30mK' system operation. This work was supported by the Czech Science Foundation under Grant No. 17-04925J. The sample preparation, characterization and high-pressure experiments were performed in MGML (http://mgml.eu/), which was supported within the program of Czech Research Infrastructures (project no. LM2018096).